\documentclass[aps,prb,reprint,superscriptaddress]{revtex4-1}
\usepackage{amsmath}
\usepackage{amsfonts}
\usepackage{graphicx}
\usepackage{color}
\newcommand{\ket}[1]
{
	\left| #1 \right>
}

\newcommand{\rket}[1]
{
	\left| #1 \right)
}

\newcommand{\rbraket}[2]
{
	\left( #1 \right|\left. #2 \right)
}
\newcommand{\ld}
{
	\,\mathrm{d}
}

\begin{document}

\title{Dynamical magnetic anisotropy and quantum phase transitions in a vibrating spin--1 molecular junction}
\author{David A. Ruiz--Tijerina}
\affiliation{Department of Physics and Astronomy and Nanoscale and Quantum Phenomena Institute, Ohio University; Athens, Ohio 45701--2979, USA}

\author{Pablo S. Cornaglia}\author{C. A. Balseiro}
\affiliation{Centro At{\'o}mico Bariloche and Instituto Balseiro, CNEA, 8400 Bariloche, Argentina} 
\affiliation{Consejo Nacional de Investigaciones Cient\'{\i}ficas y T\'ecnicas (CONICET), Argentina}

\author{Sergio E. Ulloa} \affiliation{Department of Physics and Astronomy and Nanoscale and Quantum Phenomena Institute, Ohio University; Athens, Ohio 45701--2979, USA}

\date{\today}

\begin{abstract}

We study the electronic transport through a spin--1 molecule in which mechanical stretching produces a magnetic anisotropy. In this type of device, a vibron mode along the stretching axis will couple naturally to the molecular spin. We consider a single molecular vibrational mode and find that the electron-vibron interaction induces an effective correction to the magnetic anisotropy that shifts the ground state of the device toward a non--Fermi liquid phase. A transition into a Fermi liquid phase could then be achieved, by means of mechanical stretching, passing through an underscreened spin--1 Kondo regime. We present numerical renormalization group results for the differential conductance, the spectral density, and the magnetic susceptibility across the transition. 
\end{abstract}


\maketitle

\section{Introduction}
The electronic properties of nanostructures depend critically on their symmetries. As a consequence, the ability to modify these microscopic symmetries makes it possible to drive the system through different physical regimes at will, possibly resulting in quantum phase transitions (QPTs) as the system visits different ground states.\cite{sachdev_qpt}

Quantum dots\cite{logan, Florens2011, PustilnikGH2003} and molecular devices consisting of complex molecules deposited on metallic break junctions\cite{roch_nature_2008, parks_science_2010, yu_natelson_prl_2005, scott_natelson_acsnano_2010} are good examples of systems where physical regimes can be explored by tuning their parameters. The electronic transport through molecular devices is very sensitive to the hybridization of the molecular energy levels with the bands of the metallic leads in the break junction, as well as to electron--electron (e-e) and electron--vibron interactions within the molecule. \cite{PhysRevB.69.245302,PhysRevB.68.205324,PhysRevB.68.205323,Cornaglia2004,Paaske2005,PhysRevB.72.041301,Cornaglia2005,*Cornaglia2005b,*PhysRevB.76.241403,balseiro:235409,mravlje2005,PhysRevB.76.241403,PhysRevB.79.155302}

Success has been achieved in the past few years in controlling the properties of molecular junctions through tuning of the molecular levels by different means. It has been shown, for example, that a gate voltage is capable of inducing a singlet--triplet transition in a $\text{C}_{60}$ (buckyball) molecule trapped between metallic leads, due to distinct hybridizations of these states with the electronic states of the leads.\cite{roch_nature_2008} 

In recent experiments by Parks {\it et al.},\cite{parks_science_2010} it was shown that mechanical stretching of the spin--1 molecule 
$\text{Co(tpy-SH)}_2$ (4'-mercapto-2, 2':6',2"-terpyridine) along the transport axis in a break junction setup, can be used to control the magnetic properties of the molecule. 
The stretching induces a splitting of the spin--1 triplet ground state,\cite{kroll,payen,cornaglia_epl_2011,Procolo2010} raising the energy of the doublet with spin projection $S_z = \pm 1$ by as much as 4 meV with respect to the state of $S_z = 0$,\cite{parks_science_2010} and leaving the latter as the molecular ground state. 

This magnetic anisotropy is critical to the low--temperature transport through the molecular junction in this experiment. In the absence of anisotropy, the system exhibits an underscreened spin--1 Kondo effect, signaled by enhanced conductivity for temperatures below the Kondo temperature $T_K^0$.\cite{PhysRevB.75.245329} Any positive (hard--axis) anisotropy breaks the ground state degeneracy of the isolated molecule and drives the system into a Fermi liquid ground state with an associated low conductance. \cite{zitko_prb_2008,*Zitko2010,PhysRevB.71.075305}

The converse case of negative (easy--axis) anisotropy could be reached, in principle, by compressing the molecule. This situation would set the $S_z = \pm 1$ doublet as the molecular ground state, yielding non--Fermi--liquid (NFL) behavior.\cite{Schlottmann2000,Schlottmann2001,Hewson2005Sing,zitko_prb_2008}

In this context, visiting different ground states becomes a matter of adjusting the magnetic anisotropy. The system will go from an ordinary Fermi liquid (FL) in the hard-axis regime to a NFL in the easy-axis regime, passing through an underscreened Kondo ground state\cite{Hewson2005Sing} ---a singular Fermi liquid (SFL)--- when the triplet is exactly degenerate.\cite{PhysRevB.72.014430} These different regimes and the associated QPTs can be studied as function of a single parameter: the magnetic anisotropy.

All of the effects described above can be caused by static deformations of the molecule. Moreover, it is to be expected that dynamical effects may arise via an analogous coupling between the spin and the mechanical degrees of freedom of the molecule.\cite{May2011} 
As we show below, a coupling between molecular vibrations and spin will induce a deformation in the molecular ground state, opening access to an easy--axis regime. With this motivation, in this paper we study a model that encompasses the anisotropy regimes described above, and considers in addition the mechanical degrees of freedom of the molecule through a vibrational mode. In the same way as the static deformation, vibrations along the axis couple naturally to the spin projection of the molecule. 
\begin{figure}[h]
 \includegraphics[scale=0.25]{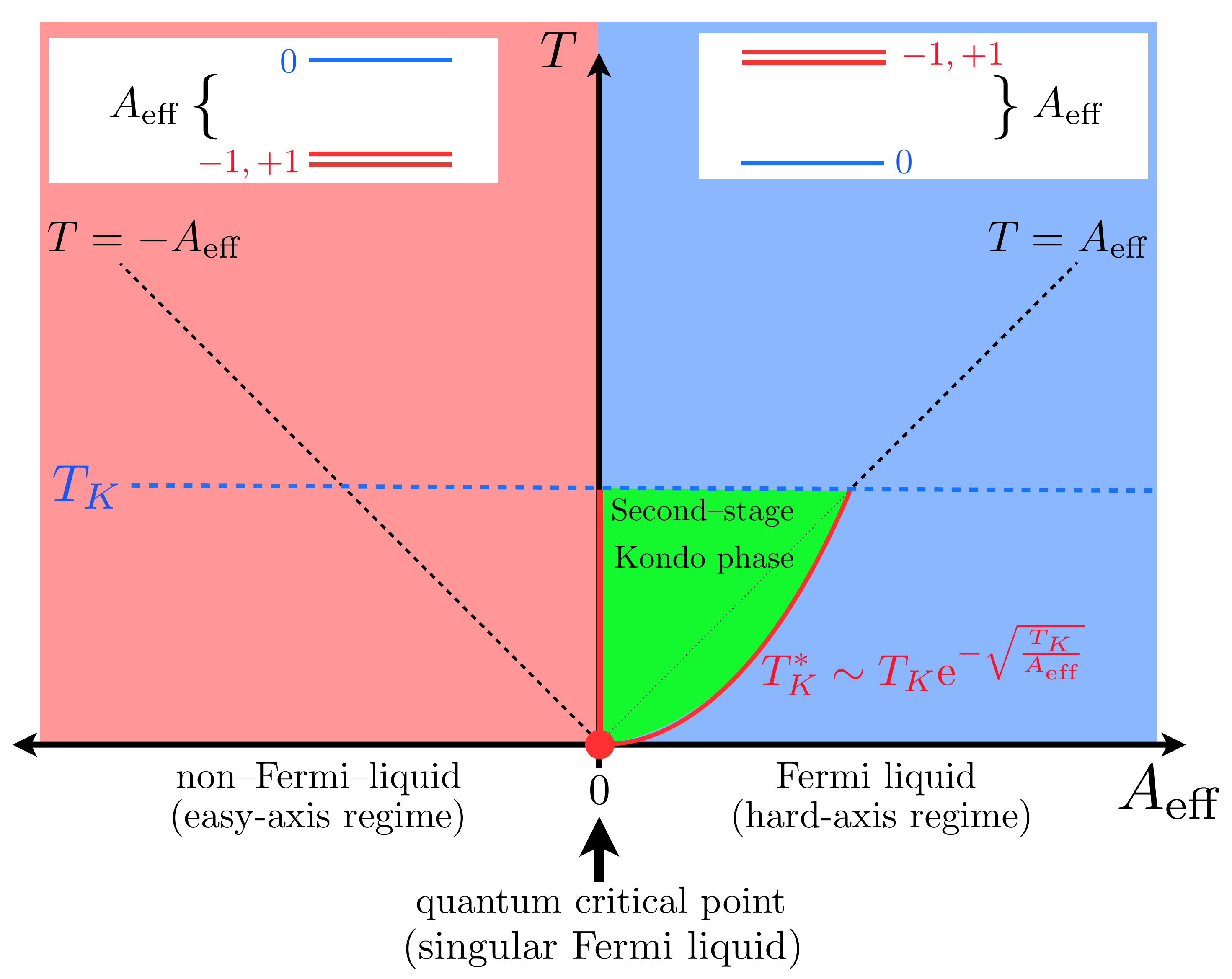}
 \caption{(Color online) Phase diagram of a deformable spin--1 molecular break junction in the presence of a local vibrational mode. A rich variety of phases can be found at temperatures below $T_K$, at which Kondo physics dominates the transport properties of the system. The effective anisotropy $A_{\text{eff}}$, induced by both static and dynamic deformations of the molecule, drives the system through different quantum phases.}
 \label{fig:phase_diagram}
 \end{figure}
In the case of an isolated molecule, we find that the coupling to the vibrational mode indeed opens access to the easy--axis regime, i.e., the spin--1 doublet of $S_z = \pm1$ becomes the ground state. 
When the molecule is coupled to leads, we find that the magnetic anisotropy is further renormalized. The resulting effective anisotropy could be tuned to explore a variety of ground states which we will discuss below.
We studied the different highly--correlated regimes of the model by means of numerical renormalization group (NRG) calculations.\cite{RevModPhys.80.395, krishnamurthy1_1980, wilson_1975} 
The results are summarized in the phase diagram shown in Fig.\ \ref{fig:phase_diagram}, to which we will come back later.

\section{Model}\label{sec:model}
\begin{figure}[h]
 \includegraphics[scale=0.245]{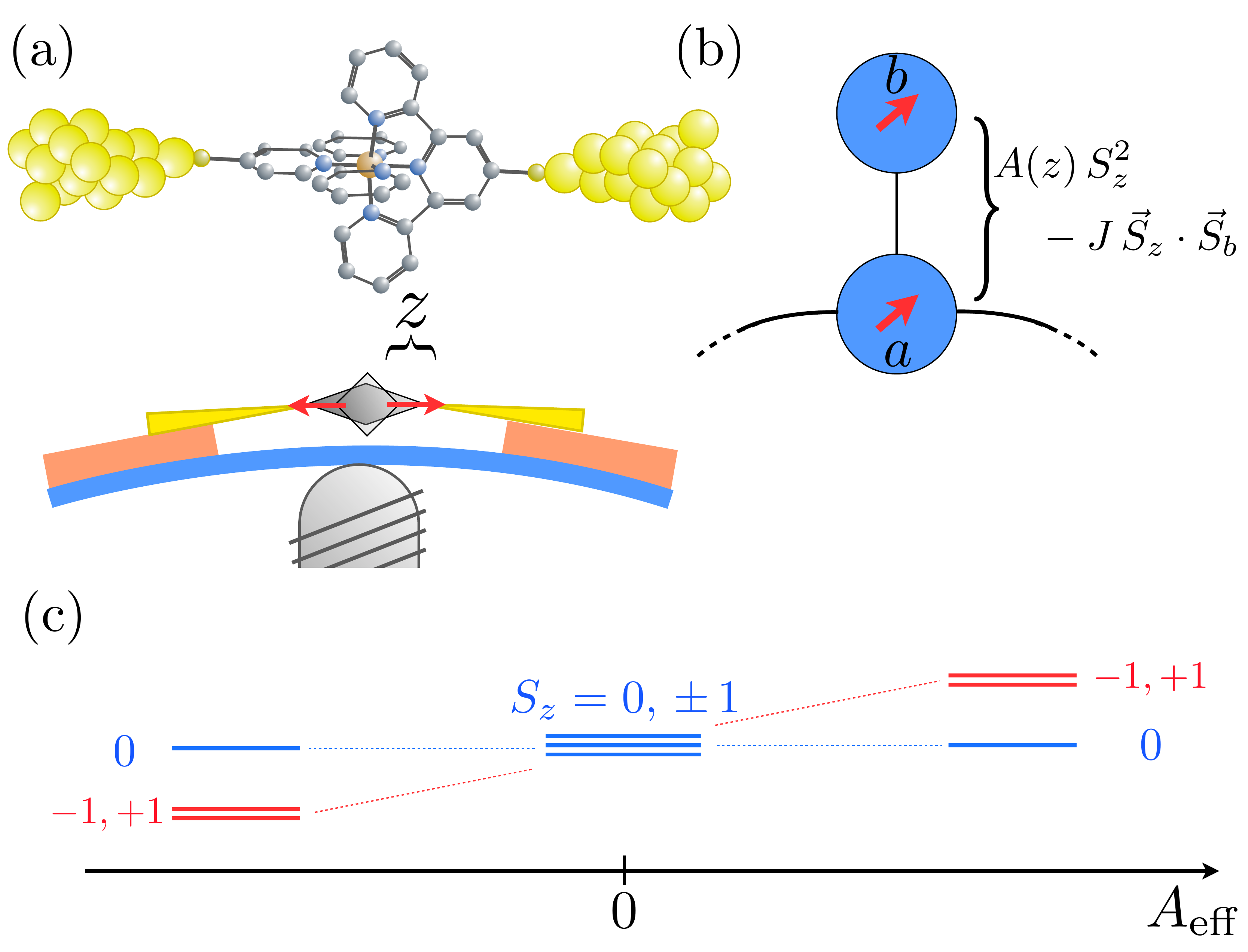}
 \caption{(Color online) Spin--1, deformable molecular device. (a) A break junction supports a spin--1 molecule that can be stretched by mechanical means. (b) The molecule is modeled by two coupled orbitals, only one of them connected to metallic leads. The spin--1 regime is enforced by an inter--orbital ferromagnetic coupling. Magnetic anisotropy, induced by static and dynamic (vibrational) stretching of the molecule, is accounted for by the term $A(z)S_z^2$ in the Hamiltonian. (c) Lowest lying level energies of the isolated molecule, obtained for different values of the effective static anisotropy.} 
 \label{fig:device}
 \end{figure}

We model the molecular device shown in Fig.\  \ref{fig:device}(a) as depicted in \ref{fig:device}(b). The Hamiltonian of the spin--1 molecule can be represented by a 2--orbital model of the form
\begin{equation}\label{eq:HD}
	H_{0} = \sum_{i=\text{a, b}}\left( \varepsilon\,n_{i} + U\,n_{i\uparrow}n_{i\downarrow} \right) - J\,\vec{S}_a\cdot \vec{S}_b.
\end{equation}
Here, $i=a,b$ are the two degenerate molecular orbitals with energy $\varepsilon$ and intra-orbital Coulomb repulsion $U$, $n_{i\sigma} = d_{i\sigma}^{\dagger}d_{i\sigma}$  and $d_{i\sigma}^{\dagger}$($d_{i\sigma}$) the creation (annihilation) operator of the corresponding orbital; $n_{i} = n_{i\uparrow} + n_{i\downarrow}$, and $\vec{S}_{i}$ is the spin operator asociated to orbital $i$. 
For simplicity, we consider the electron--hole (e-h) symmetric case where $\varepsilon = -U/2$ and the Fermi level of the leads is $\varepsilon_F=0$. The ferromagnetic ($J>0$) coupling enforces Hund's rule, setting the spin--1 triplet as the ground state of the molecule. These states are defined as
\begin{equation}\label{eq:triplet_and_singlet}
\begin{split}
	\ket{T,\,+1} = \ket{\uparrow_a\,\uparrow_b}\quad, \quad\ket{T,\,-1} = \ket{\downarrow_a\,\downarrow_b}&,\\
	\ket{T,\,0} = \frac{1}{\sqrt{2}}\Big( \ket{\uparrow_a\,\downarrow_b} + \ket{\downarrow_a\,\uparrow_b} \Big)&, \\
\end{split}
\end{equation}
where $\ket{\sigma_a\,\sigma_b}$ are the states with one electron on each orbital.
 
The deformation--induced anisotropy arising from the breaking of the octahedral symmetry can be written as a function of the elongation along the $\hat{z}$ axis, as $A(z)S_z^2 $. A purely static model [$A(z)=A_0$] of this kind was applied in reference [\onlinecite{cornaglia_epl_2011}] to describe the system of reference [\onlinecite{parks_science_2010}]. For small oscillations around the equilibrium position $z_0$ we take $A(z)= (A_0 + A_1\,\delta z)\,S_z^2$, where $\delta z=z-z_0$. 
The coupling $A_0$ represents the static deformation, whereas $A_1$ arises due to the e-ph coupling. We choose units such that $\delta z=a+a^{\dagger}$, where $a$ and $a^\dagger$ are phonon operators.  The Hamiltonian of the molecule including the electron--vibron interaction becomes
\begin{equation}\label{eq:HM}
	H_{\text{M}} = H_0 +  A_0\,S_z^2 + A_1\,S_z^2\left(a + a^{\dagger} \right) + \omega_0\,a^{\dagger}a,
\end{equation}
with $\omega_0$ the phonon frequency. By means of the unitary transformation
\begin{equation}\label{eq:transformationop}
	\tilde{H}_{\text{M}}=\mathcal{U}\,H_{\text{M}}\,\mathcal{U}^{\dagger} \quad;\quad \mathcal{U} = \exp{\left\{ -\frac{A_1}{\omega_0}S_z^2\left(a - a^{\dagger} \right) \right\}},
\end{equation}
we diagonalize (\ref{eq:HM}), which adopts the form
\begin{equation}\label{eq:HM_tilde}
	\tilde{H}_{\text{M}} = H_0 + \tilde{A} S_z^2 + \omega_0\,b^{\dagger}b,
\end{equation}
where $b=a+(A_1/\omega_0)\, S_z^2$ is a displaced phonon operator. 
Note that the magnetic anisotropy of the isolated molecule
\begin{equation}\label{eq:a_eff}
	\tilde{A} = A_0 - A_1^2/\omega_0,
\end{equation}
is negative in the absence of a static distortion ($A_0=0$).
The lowest energy eigenstates of (\ref{eq:HM_tilde}) for each molecular spin projection are
\begin{equation}\label{eq:triplet_w_phonons}
	\ket{T,\,\pm1;\,\tilde{0}} = \ket{T,\,\pm1}\rket{\tilde{0}},\quad \ket{T,\,0;\,0} = \ket{T,\,0}\rket{0},
\end{equation}
where $\rket{n}$ is an eigenvector of the operator $a^\dagger a$, and $\rket{\tilde{n}}=\text{e}^{\frac{A_1}{\omega_0}(a - a^{\dagger})}\rket{n}$ is an eigenvector of  $b^\dagger b$, with eigenvalue $n$.

We now leave the isolated molecule (molecular limit) to explore the consequences of coupling the metallic leads to the molecular orbital $a$, through the hybridization term
\begin{equation}\label{eq:hyb_term}
	H_{\text{M-E}} = \sqrt{2}\,V\sum_{\vec{k},\sigma}\left( d_{a\sigma}^{\dagger}c_{\vec{k}\sigma} + \text{H.c.} \right).
\end{equation}
This ``hanging--level'' arrangement correctly describes the low--energy behavior of the system, and other configurations can be related to it by means of a level rotation, so there is no loss of generality.\cite{cornaglia_epl_2011}
 Because we assume identical right (R) and left (L) leads and couplings, we have defined the lead--symmetric operators
\begin{equation}
	c_{\vec{k}\sigma} \equiv \left( c_{\text{L}\vec{k}\sigma} + c_{\text{R}\vec{k}\sigma} \right)/\sqrt{2},
\end{equation}
which are the only combinations that couple to the molecule. Their anti--symmetric counterparts contribute only a constant energy term to the Hamiltonian.

In the displaced basis the full transformed Hamiltonian reads
\begin{equation}\label{eq:Htilde}
\begin{split}
	\tilde{H} = H_0 &+ \tilde{A}\,S_z^2 + \omega_0\,b^{\dagger}b \\
	&+ \sum_{\vec{k},\sigma}\left( \tilde{V}_{\sigma}\, c_{\vec{k}\sigma}^{\dagger}d_{a\sigma} + \text{H.c.} \right),
\end{split}
\end{equation}
with a hybridization operator
\begin{equation}\label{eq:renormalized_v}
	\tilde{V}_{\sigma} = \sqrt{2}V\,\exp\left\{\frac{A_1}{4\omega_0}\left(a - a^{\dagger} \right)\left[1 - 2n_{a\bar{\sigma}} +4\sigma S_z^b \right]\right\}.
\end{equation}
Although the expression for the transformed Hamiltonain is more complicated than the starting one, we can appreciate an interesting feature. Namely, that it couples the $\pm1$ and $0$ spin projections of the triplet to the electronic states of the band with different strengths. As we will see below, this asymmetry leads to a contribution to the total magnetic anisotropy.
\subsection{Analytical results}
To expose the splitting of the triplet due to the spin dependent couplings to the band, we perform a Schrieffer--Wolff transformation\cite{schrieffer_wolff_PR_1966} in the limit of large $U$ and $J$. Here we set $\tilde{A}=0$ so that the isolated molecule ground state is triply degenerate. We obtain an anisotropic Kondo Hamiltonian\cite{hewson_kpthf}
\begin{equation}\label{eq:Kondo_Hamiltonian}
H_{K} = A_d S_z^2 + J_K^{\parallel}s_z\,S_z + J_K^{\perp}\left( s_x\,S_x + s_y\,S_y \right),
\end{equation}
with exchange couplings
 \begin{equation} \label{eq:Kcoupl}
 	J_K^{\parallel} = \sum_{n=0}^{\infty}\frac{4V^2\left|\rbraket{\tilde{0}}{n} \right|^2}{\frac{U}{2} + \frac{J}{4} + n\omega_0},\quad J_K^{\perp} = \text{e}^{-\left(\frac{A_1}{\omega_0} \right)^2/2} J_K^0,
 \end{equation}
where $J_K^0=4V^2/(\frac{U}{2} + \frac{J}{4})$ is the Kondo coupling in the absence of e-ph coupling ($A_1=0$), and
\begin{equation}\label{eq:splitting}
	A_d = 4V^2\left(\frac{1}{\frac{U}{2} + \frac{J}{4}} - \sum_{n}\frac{\left|\rbraket{\tilde{0}}{n} \right|^2}{\frac{U}{2} + \frac{J}{4} + n\omega_0} \right) > 0,
\end{equation}
where
\begin{equation}\label{eq:phtransition_amp}
	\left| \rbraket{\tilde{0}}{n} \right|^2 = \left(\frac{A_1}{\omega_0}\right)^{2n}\frac{\text{e}^{-\left(\frac{A_1}{\omega_0} \right)^2}}{n!}.
\end{equation}

The anisotropy term $A_d$  is a consequence of the hybridization shifting the state $\ket{T,\,0;\,0}$ down in energy further than the states $\ket{T,\,\pm1;\,0}$, for any state of the band.

The e-ph interaction induces a reduction of both Kondo couplings that become anisotropic ($J_K^\perp, J_K^\parallel < J_K^0$). In the experimentaly relevant case $U\gg \omega_0$ we have $J_K^\perp< J_K^\parallel$. For weak e-ph interaction ($A_1/\omega_0\ll 1$) and $U\gg \omega_0$ we obtain:
\begin{eqnarray}
J_K^\parallel/J_K^0&\sim& 1-\frac{\omega_0}{\frac{J}{4}+\frac{U}{2}}\left(\frac{A_1}{\omega_0}\right)^2,\\
J_K^\perp/J_K^0&\sim& 1-\frac{1}{2}\left(\frac{A_1}{\omega_0}\right)^2,\\
A_d/J_K^0&\sim& \frac{\omega_0}{\frac{J}{4}+\frac{U}{2}}\left(\frac{A_1}{\omega_0}\right)^2.
\end{eqnarray}
The expresions for $J_K^\perp$ and $A_d$ are also valid in the strong e-ph coupling regime ($A_1 \gg \omega_0$), provided $A_1^2/\omega_0\ll U$, while $J_K^\perp$ is exponentially supressed [see Eq.\  (\ref{eq:Kcoupl})] making the Kondo couplings strongly anisotropic.

As a result of the correction $A_d$, the degeneracy of the triplet is broken preventing the underscreened spin-1 Kondo effect from taking place at $\tilde{A}=0$. Therefore, a shift in the static anisotropy will be needed to restore the full spin-1 degeneracy. It is also important to point out that the anisotropy of the Kondo couplings leads to effects similar to those obtained by a static magnetic anisotropy term ($A_0$). In our case of $J_K^{\parallel} > J_K^{\perp} > 0$, this anisotropy results in a shift toward the easy--axis regime. \cite{deleo2008,zitko_prb_2008}

To summarize, the exact solution of the isolated spin--1 molecule, along with a perturbative analysis to second order in the hybridization, suggest that the main effect of the electron--phonon interaction considered in this break junction setup is to reduce the Kondo couplings, and to introduce corrections to the magnetic anisotropy. The different contributions to the magnetic anisotropy result in an effective anisotropy given by
\[
A_{\text{eff}}=A_0-A_1^2/\omega_0+ A_d + \delta A,
\]
where $\delta A$ stems from the anisotropy of the Kondo couplings.

 \section{Numerical results}\label{sec:numerical_results}
We analyze the different magnetic anisotropy regimes of the system as a function of $A_0$, $A_1$ and $\omega_0$, by means of NRG calculations. In this non--perturbative method, the continuum of electronic states in the leads is logarithmically discretized according to a parameter $\Lambda > 1$. It is then mapped onto a chain of states with exponentially decaying hopping terms that is diagonalized iteratively, defining a renormalization group transformation to the low energy spectrum.\cite{krishnamurthy1_1980}

All calculations shown below were obtained with discretization factor $\Lambda = 2.5$, with no fewer than 1200 states kept, and a constant hybridization $\Gamma = \pi \left|V\right|^2/2D=0.005\,D $, where $D$ is the half--bandwidth. We choose the parameters $\varepsilon = -U/2 = -5\,\Gamma$, with ferromagnetic coupling set to $J = 0.2\,\Gamma$. Our conclusions, however, remain qualitatively unchanged for other parameters within the Kondo regime (such as $J > \Gamma$). The spectral densities were calculated following Ref. [\onlinecite{Bulla_Costi_Vollhardt_PRB2001}].

For $A_0 = A_1 = 0$, the system is expected to have an underscreened spin--1 Kondo ground state; we have verified this with NRG calculations. This regime serves as a reference for comparison with our results in other regions of parameter space.

The infinite--dimensional boson sector of the Hilbert space has to be truncated for the numerical calculation. We analyze the regime of $\omega_0 \gtrsim T_K^0$ and $A_1/\omega_0<1$, so that relevant phonon excitations include transitions from the ground state with boson component given by $\rket{\tilde{0}}$ to states with occupation $n$; these have amplitudes given by Eq.\ (\ref{eq:phtransition_amp}). This Poisson distribution peaks at $n = A_1/\omega_0<1$ and then falls to zero rapidly, suggesting a cutoff at phonon occupation $n \sim 1$.\footnote{Within the explored range of $\omega_0$ and $A_1$, we found that a maximum occupation of 3 phonons in our NRG runs correctly describes the behavior of our system. This was explicitly verified in our numerical results, and is in accordance with the analysis of [{\protect \onlinecite{hewson_jphyscondensmatter_2002}}].}
   \begin{figure}[h]
 \includegraphics[scale=0.3]{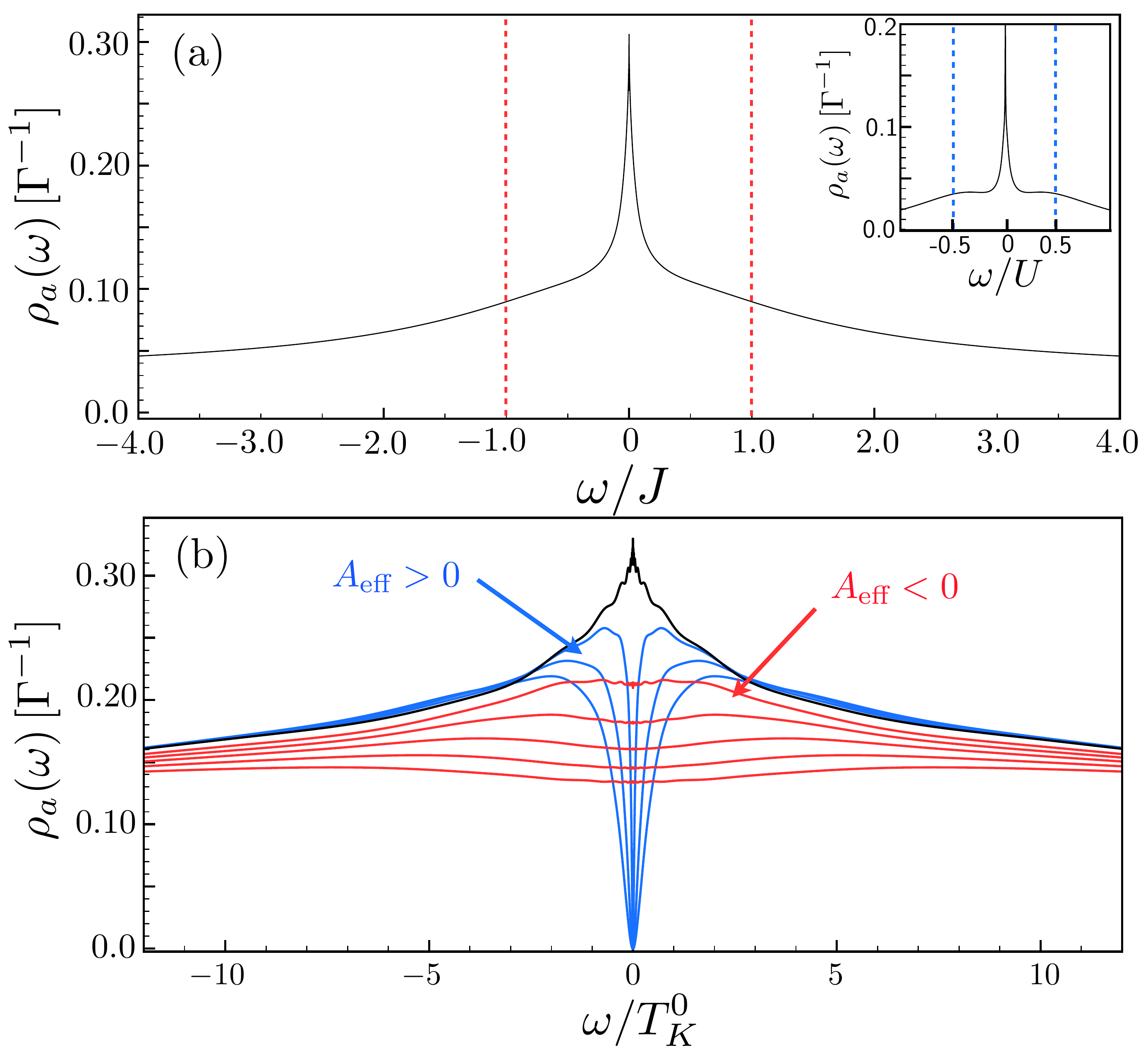}
 \caption{(Color online) (a) Spectral density at orbital $a$ for $A_0 = A_1 = 0$ (reference system). The screening of the molecular triplet ground state by the band is characterized by the sharp Kondo resonance of width $T_K^0$ at the Fermi level ($\omega = 0$), followed by shoulders at approximately the singlet--triplet (red dashed lines), and the charge (blue dashed lines in the inset) excitation energies of the isolated molecule. The Kondo effect yields unitary conductance at low ($\ll T_K^0$) temperatures. (b) Same as (a), but for finite magnetic anisotropy ($A_0 = \omega_0 = 5\,T_K^0$, $A_1$ from $2\,T_K^0$ to $11\,T_K^0$ ). In the easy axis regime of $A_{\text{eff}} < 0$ (red curves), the system presents a finite amplitude of the spectral density at the Fermi level. The hard--axis regime ($A_{\text{eff}} > 0$; blue curves) presents a dip at the Fermi level, keeping the system in a Coulomb blockade. The underscreened Kondo state is recovered when $A_{\text{eff}}=0$ (black curve).} 
 \label{fig:spectral_densities}
 \end{figure}
\subsection{Transport properties at low temperatures}
The low--temperature, linear transport properties of the system can be obtained in terms of the spectral density at orbital $a$, $\rho_a(\omega)=\sum_{\sigma}\rho_a^{\sigma}(\omega)$, which relates to the conductance through\cite{jauho_meir_wingreen_PRB1994,Pastawski1992}
 \begin{equation}\label{eq:meir_and_wingreen}
 	\frac{G(T)}{G_0} =  \pi\Gamma\int_{-\infty}^{\infty} \ld \omega \left(-\frac{\partial f(\omega,T)}{\partial \omega} \right)\,\rho_{a}(\omega),
 \end{equation}
where $G_0 = 2e^2/h$, and $f(\omega,\, T)$ is the Fermi distribution. In the reference system, where $A_0 = A_1 = 0$, $\rho_a(\omega)$ shows a sharp Kondo resonance of width $T_K^0 \sim 10^{-5}\,D$ at the Fermi level\cite{PhysRevB.75.245329, PhysRevLett.94.036802} ($\varepsilon_F$, set here to zero) as shown in Fig.\  \ref{fig:spectral_densities}(a). At low temperatures the conductance reaches the unitary limit. Shoulders in the reference spectral density appear at approximately the singlet--triplet excitation energy, $\omega \sim \pm J$, and peaks at the charge excitation energies $\omega \sim \pm U/2$.

With this in mind we move on to analyze Fig.\  \ref{fig:spectral_densities}(b) in the hard--axis regime ($A_{\text{eff}} > 0$). The spectral density has two peaks close to the Fermi level separated by a sharp dip, which according to (\ref{eq:meir_and_wingreen}) puts the device in a transport blockade with a vanishing zero--bias conductance at $T \rightarrow 0$. 

In the easy--axis regime ($A_{\text{eff}} < 0$), conduction electrons will scatter off the molecule's ground state,\cite{zitko_prb_2008} the doublet $\ket{T,\pm1}$.  At zero temperature $\rho_a(\varepsilon_F)$ presents a non--zero amplitude, which means that the system will conduct at zero bias, although not unitarily as in the case of zero anisotropy.
\begin{figure}[h]
 \includegraphics[scale=0.25]{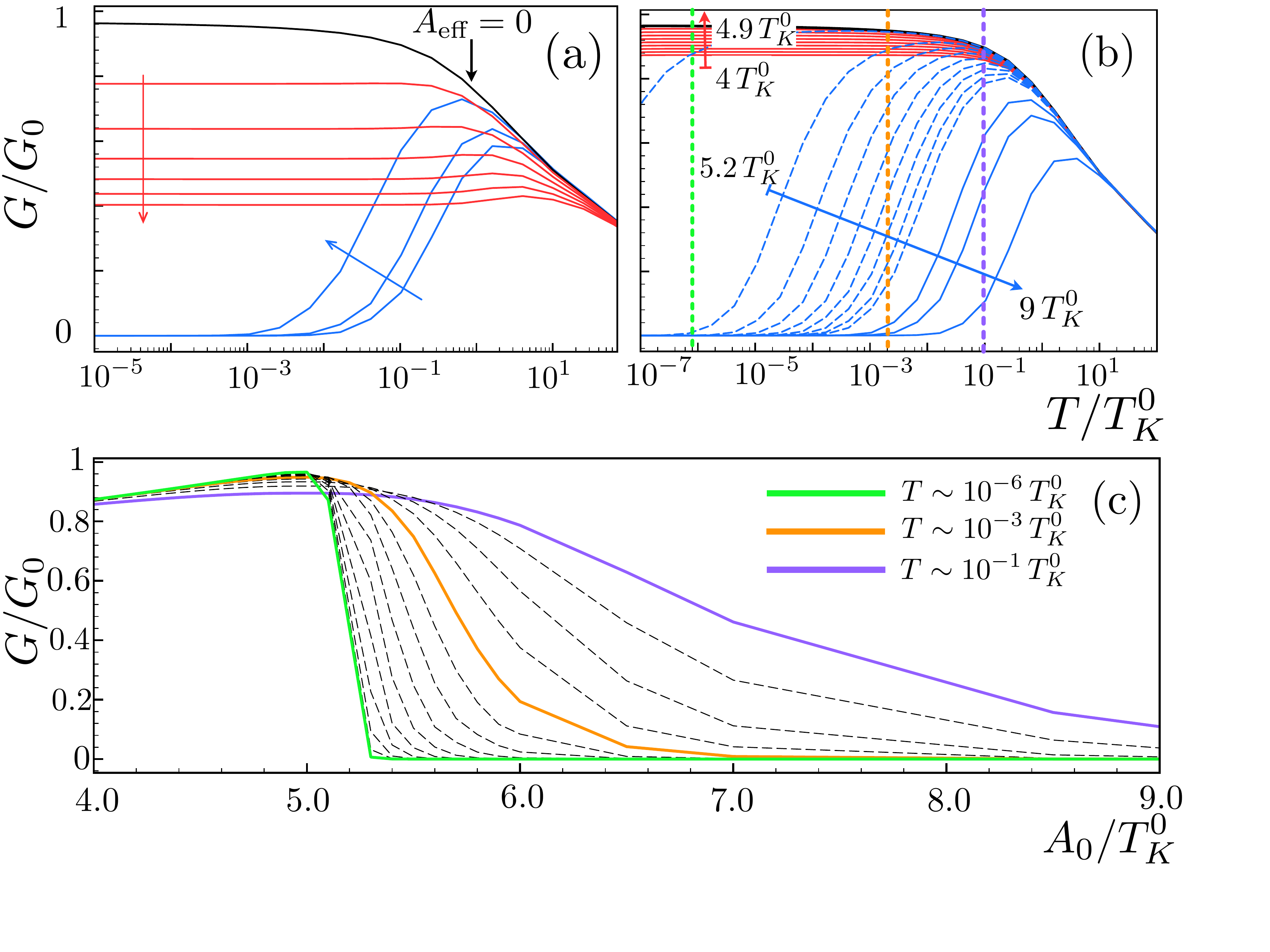}
 \caption{(Color online) Conductance as a function of temperature, for different values of $A_{\text{eff}}$. In (a), varying $A_1$ from $2\,T_K^0$ to $12\,T_K^0$ for $\omega_0 = \,A_0 = 5\,T_K^0$, and in (b) by varying $A_0$ from $4\,T_K^0$ to $9\,T_K^0$ for $A_1 = \omega_0 = 5\,T_K^0$. In (c), as a function of $A_0$ for different isotherms. The zero--temperature conductance vanishes for positive anisotropy, due to the lack of available states for transport in the molecule at the Fermi level (see Fig.\  \ref{fig:spectral_densities}). As $A_{\text{eff}} \rightarrow 0$ the conductance curves vary as indicated by the blue arrows until the unitary limit is reached, signaled in the plot by a plateau that extends to lower temperatures. Unitarity is then lost due to many--body scattering of the leads' electrons off the $S_z = \pm1$ degenerate ground state of the molecule at negative anisotropies. As $A_{\text{eff}}$ goes from zero to negative the conductance curves follow the trend indicated by the red arrow. 
In (c) we show isotherms for the conductance as a function of $A_0$ (corresponding to the curves in (b)), which can be directly related to experiments in which the conductance is measured in a stretchable device, at constant temperature. The quantum critical point becomes apparent at very low temperatures, where a sudden drop in conductance signals the change of sign of $A_{\text{eff}}$. The Kondo temperatures of these cases are lower than the reference scale $T_K^0$ by one order of magnitude, as can be seen by the onset of the conductance plateaus below $T = T_K^0$.}
 \label{fig:conductances}
 \end{figure}
The different regimes described above are shown in Fig.\  \ref{fig:spectral_densities}(b) from NRG calculations in which $A_{\text{eff}}$ was tuned by varying $A_1$, at constant static anisotropy $A_0$ and phonon frequency $\omega_0$.

Curves of conductance as a function of temperature, corresponding to the parameters of Fig.\  \ref{fig:spectral_densities}(b), are shown in Fig.\  \ref{fig:conductances}(a), whereas in Fig.\  \ref{fig:conductances}(b) the same regimes are explored by varying $A_0$ instead. As mentioned above, tuning $A_0$ in molecular devices has been achieved experimentally by Parks \emph{et al}.,\cite{parks_science_2010} through stretching. Assuming that the value of $A_0$ increases from zero by stretching the molecule, it could be used to change the sign of the effective magnetic anisotropy. This is depicted in Fig.\  \ref{fig:conductances}(b), where we begin with a positive value of $A_0$ such that the effective anisotropy is negative, and then increase $A_0$ until we reach the hard--axis regime. Experiments such as those of reference [\onlinecite{parks_science_2010}] may perhaps be more easily related to Fig.\  \ref{fig:conductances}(c), where we show the zero--bias conductance at constant temperature, for different values of $A_0$. The sudden drop in conductance is a clear signature of the transition from hard--axis to easy--axis behavior at zero temperature.

\subsection{Effective anisotropy and underscreened Kondo effect restoration}
\begin{figure}[h]
 \includegraphics[scale=0.37]{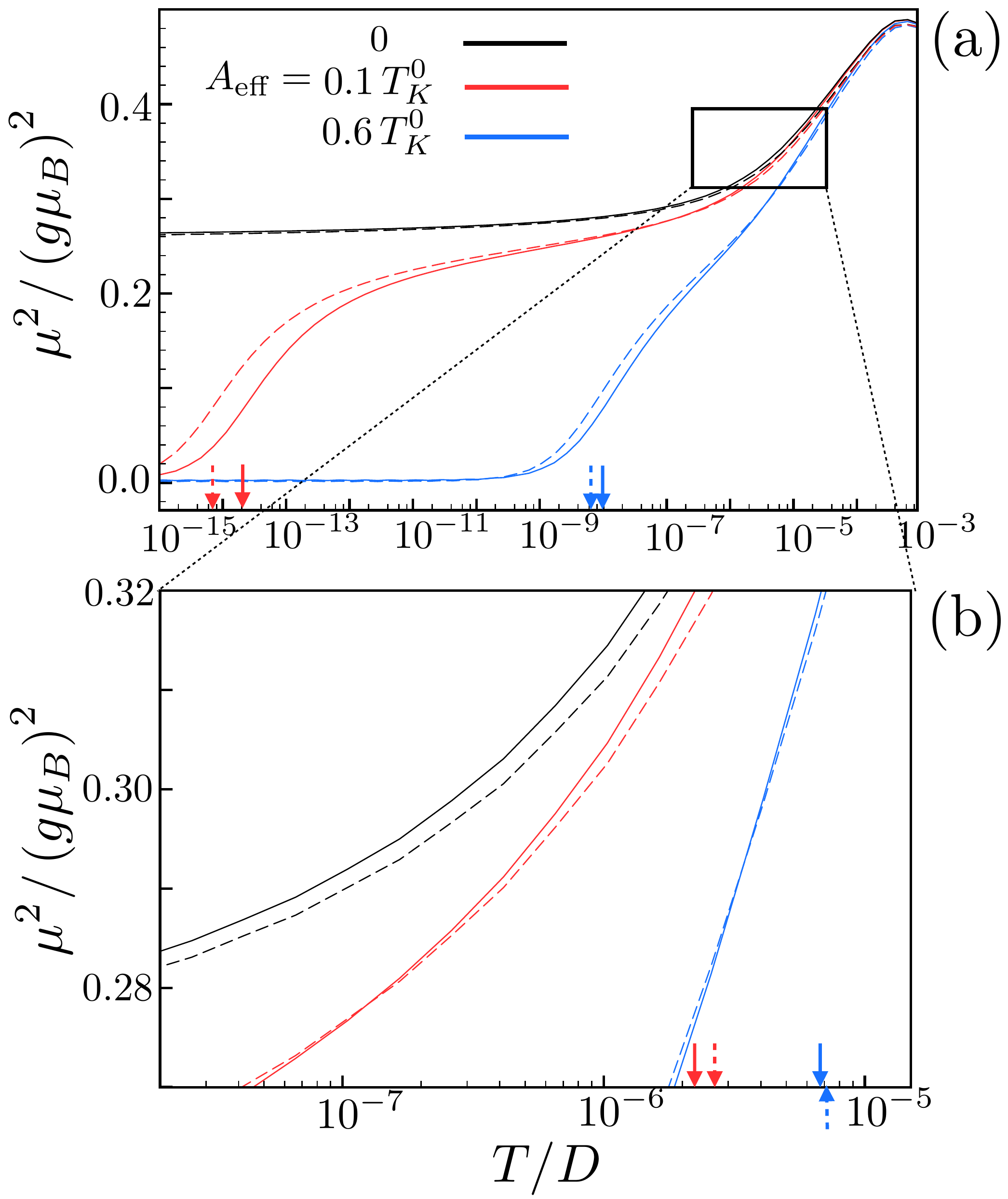}
 \caption{(Color online) Effective magnetic--moment--squared, $\mu^2 \equiv T\,\chi(T)$, as a function of temperature for different values of the magnetic anisotropy. Solid lines are for a system with e-ph interaction ($A_1\neq 0$), whereas dashed lines correspond to a system without e-ph interaction, and with a static magnetic anisotropy tuned to match the effective anisotropy of the e--ph interacting system. As expected from the e--ph mediated reduction of the Kondo exchange couplings, the first--stage Kondo temperature is lower in the e-ph interacting system [solid arrows in (b)], that is, $T_K < T_K^0$. This in turn \emph{increases} the second--stage Kondo temperature $T_K^*$ [arrows in (a)], as expected from Eq.\  (\ref{eq:second_kondo}). 
} 
 \label{fig:comparison}
 \end{figure}
As we will see below, the numerical results confirm that the main effect of the e--ph coupling considered is to renormalize the magnetic anisotropy toward the hard--axis regime and reduce the Kondo exchange couplings. In fact, the static anisotropy term can be tuned to recover the underscreened spin--1 Kondo effect.
We define $\Delta= A_{\text{eff}}-\tilde{A}$ and calculate it numerically through an analysis of the effective magnetic--moment--squared, $\mu^2$, at low temperature.\cite{wilson_1975} The onset of Kondo screening is signaled by a drop in the molecule's magnetic moment at temperatures below some $T_K(A_0,\, A_1,\, \omega_0)$. Notice that $T_K < T_K^0$ due to the reduced Kondo couplings [see Eqs. (\ref{eq:Kcoupl})]. An example of this is shown in Fig.\  \ref{fig:comparison}(b).

As the molecule is coupled to a single conduction electron channel, the electrons in the leads are able to screen one half of the molecule's spin, leaving an asymptotically free spin--$1/2$ object\cite{Hewson2005Sing, nozieres-paris_1980} and producing a plateau $\mu^2(T<T_K) = 0.25$, as in Fig.\  \ref{fig:comparison}. Further screening associated to a second--stage Kondo effect is known to arise when a net positive anisotropy $0< A_{\text{eff}}\ll T_K$ is present. We have $\mu^2(T) \sim 0$ for temperatures below a second stage Kondo temperature $T_K^*$, given by\cite{cornaglia_epl_2011}
\begin{equation}\label{eq:second_kondo}
	T_K^* = c_1\,T_K\,\text{e}^{-2\,c_2\sqrt{\frac{T_K}{A_{\text{eff}}}}},
\end{equation}
with constants $c_1,\,c_2 \sim 1$ that depend weakly on the parameters.

On the easy--axis side, the Kondo screening is suppressed by the energy gap between the molecular ground state $\ket{T,\,\pm1;\,0}$ and the first excited state $\ket{T,\,0;\,0}$. In other words, Kondo screening becomes increasingly less efficient as we go farther away from full degeneracy of the triplet, and we obtain $0.25 < \mu^2 < 1$ for all temperatures (not shown).
 
\begin{figure}[h]
 \includegraphics[scale=0.27]{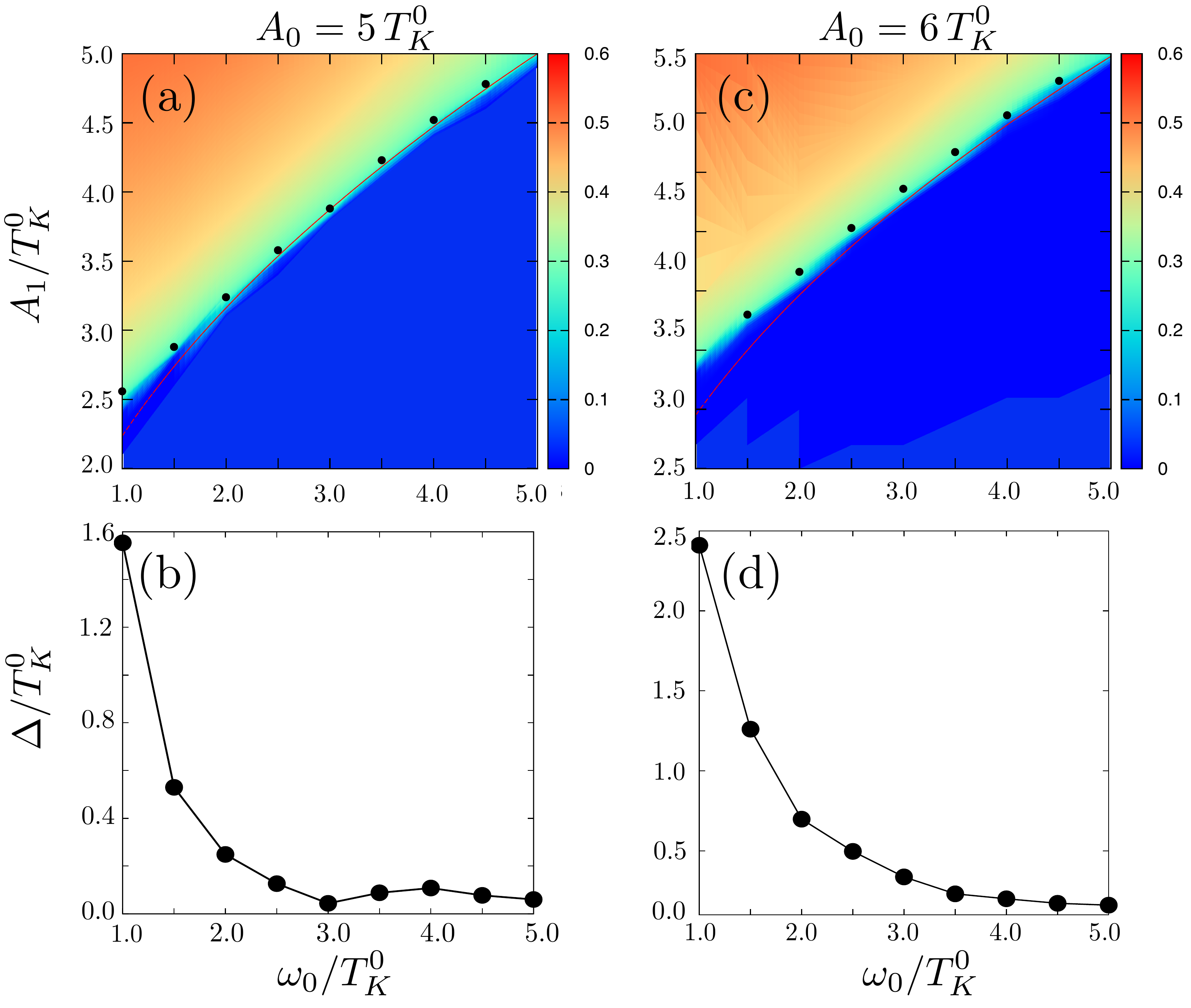}
	 \caption{(Color online) (a) and (c): Magnetic--moment--squared of the molecule at zero temperature ($\mu^2(T\rightarrow0)$) as a function of $A_1$ and $\omega_0$, for (a) $A_0 = 5\,T_K^0$ and (c) $A_0 = 6\,T_K^0$. In each case, the red curve ($A_1 = \sqrt{\omega_0A_0}$) indicates the parameters of full degeneracy of the spin--1 triplet in the molecular limit. The black dots indicate the parameters where the underscreened spin--1 Kondo effect is recovered ($\mu^2 = 0.25$). (b) and (d): Correction to the magnetic anisotropy due to the coupling to the leads for the parameters of (a) and (c), respectively. } 
 \label{fig:colormap}
 \end{figure}

In order to quantify $A_{\text{eff}}$ and estimate $\Delta$, we carry out NRG calculations at fixed $A_0$, varying $A_1$ and $\omega_0$. The transition from a hard to an easy axis is observed as a sudden jump (at zero temperature) from the hard--axis value of $\mu^2 = 0$ to $\mu^2 = 0.25$ in the spin--1 Kondo regime, as shown in Fig.\  \ref{fig:colormap}(a).

For the molecular limit where the magnetic anisotropy coefficient is exactly given by $\tilde{A}$, the change of regime is indicated by the red curve in Fig.\  \ref{fig:colormap}(a) and (c), given by $\tilde{A}=0$. When the coupling to the band is taken into account, the transition occurs for $A_{\text{eff}} = 0$ and is signaled by $\mu^2(T\rightarrow 0) = 0.25$. We indicate this transition with black dots in the figures, and find $\Delta=A_{\text{eff}}- \tilde{A}>0$ as expected from our discussion above. The values of $\Delta$ for every $\omega_0$ are presented in Fig.\  \ref{fig:colormap}(b) and (d) for their respective maps. Notice that the points at larger $\omega_0$ correspond to smaller $A_1/\omega_0$, which explains the decreasing trend of $\Delta$ with increasing $\omega_0$.

The e--ph interactions suppress the Kondo couplings, reducing the spin--1 Kondo temperature $T_K$ of the device. This results, according to Eq.\  (\ref{eq:second_kondo}), in an enhancement of $T_K^*$. This can be observed in Fig.\  \ref{fig:comparison}, where we compare our device to one with only static anisotropy, and find that they behave similarly, differing only in the values of $T_K$ and $T_K^*$.

\begin{figure}[h]
 \includegraphics[scale=0.28]{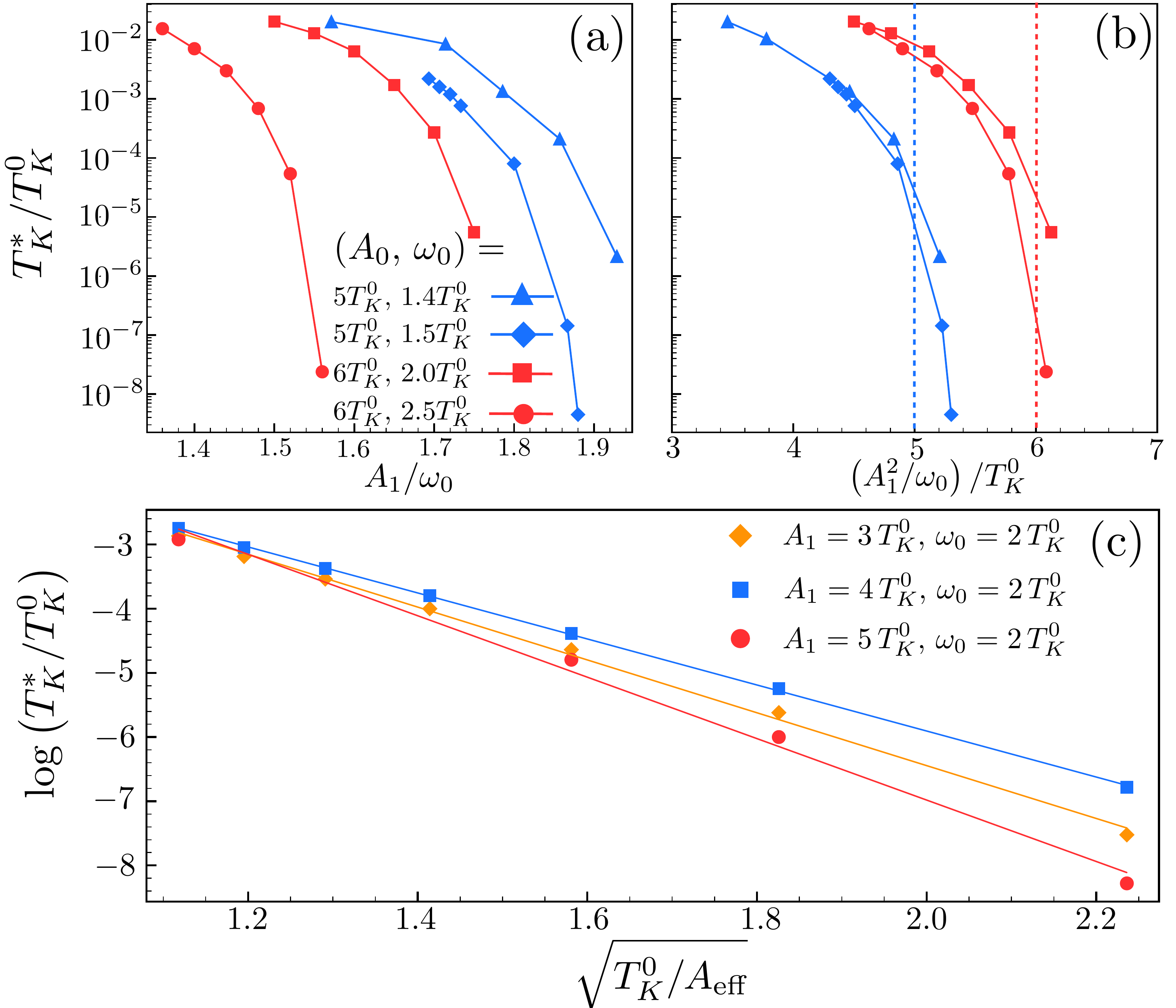}
 \caption{(Color online) Second--stage Kondo temperature, $T_K^*$, for different values of $A_1$ and $\omega_0$. (a) A fast fall of $T_K^*$ as a function of $A_1/\omega_0$ is observed, consistent with a vibronic induced shift of the magnetic anisotropy and Eq.\  (\ref{eq:second_kondo}). (b) Plotting $T_K^*$ against $A_1^2/\omega_0 = A_0 - \tilde{A}$, makes the presence of a correction $\Delta$ to the anisotropy evident; the dotted lines indicate the value of $A_1$ where the molecular limit anisotropy $\tilde{A}$ changes sign, for the corresponding values of $A_0$. The numerical results are clearly shifted toward positive values of $A_{\text{eff}}$. (c) Verification of Eq.\  (\ref{eq:second_kondo}) within our picture of effective anisotropy and suppressed Kondo exchange couplings, for fixed $A_1$ and $\omega_0$, varying $A_0$. We substitute $A_{\text{eff}}$ in the equation, with $\Delta$ computed from numerical results such as those of Fig. \ref{fig:colormap}(b) and (d). The slope of each curve is proportional to $-2\sqrt{T_K/T_K^0}$, with $T_K$ the Kondo temperature corresponding to the curve.}
 \label{fig:TK*_vs_A1}
 \end{figure}

As further verification, we utilize the values of $\Delta$ obtained from Fig.\  \ref{fig:colormap} to compute values of $A_{\text{eff}}$ for substitution into Eq.\ (\ref{eq:second_kondo}), and the degree of agreement is excellent, as can be appreciated in Fig.\ \ref{fig:TK*_vs_A1}(c).

The excellent fit to the theory reassures us of our picture, in which all the effects of the electron--phonon (e--ph) interactions can be absorbed into an effective anisotropy term, and a reduced hybridization due to polaronic effects.
 \section{Conclusions}
We have studied the behavior of a stretchable spin--1 molecule deposited on a break junction, that presents a vibrational mode along the junction axis. We performed NRG calculations to study the Kondo physics of this system; its signatures in transport and thermodynamic quantities, as well as the nature of the system's ground state. We find that the vibrational degrees of freedom induce a negative magnetic anisotropy that shifts the ground state of the undeformed molecule toward an easy--axis regime, which would be accessible to the static molecule only through compression. Polaronic corrections arising from the electron--phonon interactions suppress the Kondo exchange couplings of the molecular spin with the leads' fermionic states, in a spin--asymmetric fashion. This reduces the Kondo temperature of the device and introduces an effective correction to the magnetic anisotropy.

Static stretching of the molecule could then be used to explore a QPT from the non--Fermi--liquid ground state of the easy--axis regime, to the Fermi liquid of the hard--axis regime, visiting an underscreened Kondo critical point at zero anisotropy. The resulting phase diagram of the system is presented in Fig.\  \ref{fig:phase_diagram}. The different phases exhibit clear conductance signatures that can be explored experimentally. 

From our analysis, we may expect the un--stretched device in the experiment of Parks {\it et al.}, to be in an easy--axis regime due to the coupling to molecular vibrations.  In that case, a careful analysis of the low-temperature behavior of the conductance, such as that indicated inf Fig.\ \ref{fig:conductances}(c), would make it possible to identify the signatures of this QPT in a well-controlled environment.  

\begin{acknowledgments}
DRT and SU thank E. Vernek for helpful discussions, and especially for the NRG code that served as an important first component of this work. DRT and SU acknowledge support from NSF PIRE and CIAM/MWN (US), and CONACyT (M\'exico). PSC and CAB acknowledge financial support from PIP 1821 of CONICET and PICT-Bicentenario 2010-1060 of the ANPCyT.
\end{acknowledgments}
\bibliography{bibliography_APS.bib}{}
\bibliographystyle{apsrev4-1}
\end{document}